# Structural, electronic properties and Fermi surface of $ThCr_2Si_2$-type charge-balanced $KFe_2AsSe$ phase as a parent system for a new group of "mixed" pnictide-chalcogenide superconductors.

Igor R. Shein, * and Alexander L. Ivanovskii

*Institute of Solid State Chemistry, Ural Branch of the Russian Academy of Sciences, 620990 Ekaterinburg, Russia*

**ABSTRACT**

The $ThCr_2Si_2$-type arsenide-selenide phase $KFe_2AsSe$ is proposed as a parent system for a new "intermediate" group of Fe-based superconducting materials bridging the known families of Fe-pnictides (such as $BaFe_2As_2$) and Fe-chalcogenides (such as $K_xFe_{2-y}Se_2$) superconductors. The characterization of the proposed charge-balanced phase by means of FLAPW-GGA approach covers the crystal structure, As/Se atomic ordering, stability, electronic bands, Fermi surface, and density of electronic states.

Among the broad family of Fe-based superconductors (SCs), the ternary $ThCr_2Si_2$-type *iron-pnictide* (Fe-*Pn*) compounds belong to one of the most interesting and intensely studied groups of such materials.[1-7] Their parent phases are $AFe_2Pn_2$, where $A$ are alkali earth metals or Eu.

Very recently, the newest group of related ternary $ThCr_2Si_2$-type *iron-chalcogenide* (Fe-*Ch*) compounds was discovered.[8] Their parent phases are $A'Fe_2Ch_2$, where $A'$ are alkali metals or Tl.[8-11]

These two groups of materials (termed also 122 Fe-*Pn* and 122 Fe-*Ch* systems) adopt a quasi-two-dimensional (2D) tetragonal crystal structure (space group $I4/mmm$, #139) where [$Fe_2Pn_2$] ([$Fe_2Ch_2$]) blocks are separated by $A$ ($A'$) atomic sheets. In turn, inside [$Fe_2Pn_2$] ([$Fe_2Ch_2$]) blocks, Fe ions form a square lattice sandwiched between two $Pn$ ($Ch$) sheets shifted so that each Fe is surrounded by a distorted $Pn$ ($Ch$) tetrahedron {$FePn_4$} ({$FeCh_4$}).

However, the aforementioned groups of 122 Fe-based SCs show a set of fundamental differences. So, undoped $AFe_2Pn_2$ phases exhibit a collinear antiferromagnetic (AF) spin density wave (SDW), whereas the superconductivity emerges owing to their hole or electron doping (or external pressure).[1-7] On the contrary, for $A'Fe_2Ch_2$, the superconductivity arises in proximity to a Mott insulator. Besides, the electronic structure of $A'Fe_2Ch_2$ phases differs considerably from that of related $AFe_2Pn_2$ SCs. In particular, in contrast to $AFe_2Pn_2$ phases with multiple electron and hole bands on the Fermi surface, for $A'Fe_2Ch_2$ there are no hole pockets near the Brillouin zone center.[12-14]

Generally, these conspicuous differences reflect various electron count for $AFe_2Pn_2$ versus $A'Fe_2Ch_2$ systems, when the 122 Fe-*Ch* phases with formal compositions $A'Fe_2Ch_2$ are electron over-doped systems in comparison to $AFe_2Pn_2$ phases. Besides, taking into account the usual oxidation numbers of atoms: $A'^{1+}$, $Fe^{2+}$, and $Ch^{2-}$, the nominal formulas $A'Fe_2Ch_2$ imply the infringement of electroneutrality for these crystals, whereas the actual compositions of the synthesized 122 Fe-*Ch* phases are cation-deficient and are close to so-called charge-balanced compositions $A'_{0.8}Fe_{1.6}Ch_2$ (or $A'_2Fe_4Ch_5$).[15] The presence of a large amount of cation vacancies (both in $A'$ and Fe sites) required to maintain the charge balance for these materials, together with long-range order of Fe-vacancies,[16,17] is the major feature of 122 Fe-*Ch* phases as compared with stoichiometric $AFe_2Pn_2$.

Since $ThCr_2Si_2$-type phases show high chemical flexibility to a large variety of constituent elements,[3-7,18] we believe that partial replacement $Ch \leftrightarrow Pn$ can open up an efficient way to search for new "intermediate" group of *iron-pnictide-chalcogenide* (Fe-*Pn-Ch*) compounds *without (or with controlled) cation deficiency* bridging the aforementioned groups of 122 Fe-*Pn* and 122 Fe-*Ch* superconductors.

In this Brief Report, we focus on the charge-balanced arsenide-selenide phase $KFe_2AsSe$, which strictly obeys the electroneutrality rule, as for example the proposed "mixed" $ThCr_2Si_2$-type iron-pnictide-chalcogenide systems. The characterization of the $KFe_2AsSe$ phase by means of FLAPW-GGA approach covers the crystal structure, As/Se atomic ordering, stability, electronic bands, Fermi surface, and density of electronic states.

For the proposed $KFe_2AsSe$ system, four structural models (Fig. 1) were examined to determine how possible distributions of As and Se atoms influence the stability and properties of this "mixed" arsenide-selenide phase. For two structures (*1* and *2,* tetragonal, space groups $I4/mmm$ and $P4/mmm$, respectively), the building blocks [$AsFe_2Se$] (with alternation of atomic sheets as As/Fe/Se) are stacked in the sequence ..[$AsFe_2Se$]/[$AsFe_2Se$].. (*1*) or ..[$AsFe_2Se$]/[$SeFe_2As$]/[$AsFe_2Se$].. (*2*). Here, possible inter-blocks interactions[19,20] are only As-Se (*1*) or As-As and Se-Se (*2*). For two alternative structures (*3* and *4*, orthorhombic, space group $C/mmm$), the building blocks can be defined as [($As_{0.5}Se_{0.5}$)$Fe_2$($As_{0.5}Se_{0.5}$)]; this means that each network around a Fe sheet will consist of an equal number of As and Se atoms. In addition, two stacking types were examined, which make it possible to form Se-Se and As-As (*3*) *versus* "mixed" As-Se inter-blocks bonds (*4*), see Fig. 1.

Thus, the chosen models (*1-4*) enable us to clarify the role of As/Se atomic arrangement inside [Fe$_2$AsSe] blocks, *i.e.* [AsFe$_2$Se] *versus* [(As$_{0.5}$Se$_{0.5}$)Fe$_2$(As$_{0.5}$Se$_{0.5}$)], and the basic questions related to the types of stacking of these blocks, *i.e.* possible advantages of Se-Se and As-As (*3*) *versus* "mixed" As-Se inter-blocks interactions.

Our band-structure calculations were carried out by means of the full-potential method with mixed basis APW+lo (LAPW) implemented in the WIEN2k suite of programs.[21] The generalized gradient correction (GGA) to exchange-correlation potential in the PBE form [22] was used. The muffin-tin (MT) sphere radii were chosen to be 2.5 a.u. for K, 2.25 a.u. for Fe and 2.0 a.u. for As and Se. The plane-wave expansion was taken to $R_{MT} \times K_{MAX}$ equal to 8. The calculations were performed with full-lattice optimization including internal coordinates. The self-consistent calculations were considered to be converged when the difference in the total energy of the crystal did not exceed 0.1 mRy and the difference in the total electronic charge did not exceed 0.001 *e* as calculated at consecutive steps.

The obtained structural parameters for KFe$_2$AsSe (*1 – 4*) are presented in Table I. Since we assume KFe$_2$AsSe as a parent phase for the group of Fe-based superconductors, it seems useful to discuss these data in relation to so-called structural indicators which are used for *A*Fe$_2$*Pn*$_2$ and *A'*Fe$_2$*Ch*$_2$ SCs [10,23-25] to reflect the correlations between the structural properties *versus* $T_C$. One of them is the relation between $T_C$ and the *Pn(Ch)*-Fe-*Pn(Ch)* bond angles, *i.e.* the deviation from the angles for regular {Fe*(Pn,Ch)*$_4$} tetrahedrons ($\Theta$ = 109.47º), see Ref. [23]. Another indicator is the so-called anion height $\Delta z$, *i.e.* the distance of *Pn(Ch)* atoms from the Fe plane. Here, the value $\Delta z$ = 1.38 Å is proposed [24,25] as an "optimal" factor favorable for superconductivity in Fe-based systems. In our case, for all structures *1-4* the values of $\Delta z$ are comparable (1.33 - 1.34 Å) and are quite close to the aforementioned "optimal" factor, - as well as the bonds angles, see Table I.

Next, the differences in the calculated total energies for the examined models (*1 – 4*) are very small, < 0.05 eV/f.u. This means that the influence of As/Se distributions inside [Fe$_2$AsSe] blocks on the stability of the proposed KFe$_2$AsSe phase is very insignificant. This result can be easily explained taking into account the average inter-atomic distances $d_{As(Se)-As(Se)}$ inside [Fe$_2$AsSe] blocks for KFe$_2$AsSe (about 3.83 Å), which are comparable with the sum of the van der Waals radii of these atoms (3.75 Å,[26] and 4.00 Å [27]) and are considerably larger than the average lengths of As-Se covalent bonds: ~2.42 Å for binary phase As$_2$Se$_3$.[28]

Certainly, the strong inter-block As-As, As-Se or Se-Se interactions (where $d_{As(Se)-As(Se)} \sim$ 4.23 Å) are practically absent too, see Fig. 2, where the bonding picture for KFe$_2$AsSe (*1*) is visualized using the charge density map. Here, the formation of the strong covalent Fe-(As,Se) bonds inside [Fe$_2$AsSe] blocks is also visible.

So, we can conclude that the inter-atomic bonding for KFe$_2$AsSe is typical of other layered Fe-based superconductors [4-7] and has a highly anisotropic character, where the inter-block bonding adopts the ionic type with participation of sheets of potassium atoms. Indeed, Bader analysis [29] allows us to estimate the effective atomic charges, which demonstrate strong similarity for all structural models (*1-4*): K$^{0.77+}$, Fe$^{1.60+}$, As$^{-1.31-}$, and Se$^{1.12-}$. Thus, for KFe$_2$AsSe the charge transfer (~ 0.77 *e*) occurs from K$^{\delta+}$ sheets (which act as "charge reservoirs" [4-7]) to [Fe$_2$AsSe]$^{\delta-}$ blocks; besides, inside [Fe$_2$AsSe]$^{\delta-}$ blocks, the ionic bonding takes place between Fe-(As,Se) atoms owing to partial Fe $\rightarrow$ (As,Se) charge transfer.

Let us discuss the electronic properties of KFe$_2$AsSe using two related structures *1* and *2* as an example (Figs. 3, 4, and Table II) since the results obtained for other structures are similar.

Figure 3 shows the total and atomic-resolved *l*-projected DOSs for KFe$_2$AsSe (*1* and *2*) as calculated for equilibrium geometries. These DOSs are very similar. The valence states occupy the energy interval from $E_F$ to -6 eV, where the states from -6 eV to -4 eV and from -4 eV to -2 eV are mainly of Se 4*p* and As 4*p* character, respectively, and are hybridized with Fe 3*d* states. The near-Fermi states are mainly of the Fe 3*d* type, Table II. Note that the value of the total DOS at the Fermi level, N($E_F$), for KFe$_2$AsSe is comparable with that for KFe$_2$Se$_2$ (3.81 – 3.94 states/eV·f.u. as calculated for experimental lattice parameters [12,30]) and for $A$Fe$_2$Pn$_2$ phases, for example, for BaFe$_2$As$_2$ N(E$_F$) ~ 4.55- 4.59 states/eV·f.u. [31,32]

As the electronic bands near the Fermi surface are involved in the formation of the superconducting state, it is important to elucidate their nature. Figure 4 shows the near-Fermi bands for KFe$_2$AsSe (*1* and *2*). Note that for $A$Fe$_2$Pn$_2$ phases, the Fermi level intersects the quasi-two-dimensional (2$D$) low-dispersive Fe 3$d$ - like bands which form the characteristic Fermi surfaces (FSs) consisting of hole- (in the centre of the Brillouin zone) and electron cylindrical-like sheets (in the corners of the Brillouin zone) along the $k_z$ direction.[4-7,31,32] In turn, for electron over-doped $A'$Fe$_2$Ch$_2$, the Fermi level is shifted from the manifold of aforementioned quasi-2$D$ low-dispersive bands, which become filled, to the region of the upper bands with higher dispersion E($k$). As a result, the Fermi surfaces for $A'$Fe$_2$Ch$_2$ phases differ markedly from the FSs of $A$Fe$_2$Pn$_2$ phases and consist of two electron-like sheets in the corners of the Brillouin zone and closed disconnected electron-like pockets (around Z point). [31,32]

For the proposed charge-balanced arsenide-selenide phase KFe$_2$AsSe (*1*), an "intermediate" type of the near-Fermi band structure and the FS topology was found, Fig. 4. Indeed, two 2$D$-like bands are above E$_F$ (along $\Gamma$-Z) and form two cylindrical hole pockets (as in $A$Fe$_2$Pn$_2$ phases), two other bands are below E$_F$ and are responsible for the formation of the doubly degenerated electron pocket in the corners of the Brillouin zone along the $k_z$ direction (as in $A$Fe$_2$Pn$_2$ and $A'$Fe$_2$Ch$_2$ phases). The fifth band intersects the Fermi level along $\Gamma$-Z forming the closed disconnected electron-like pocket (as in for $A'$Fe$_2$Ch$_2$ phases), which is inserted into hole cylindrical-like sheets, see the bottom panel in Fig. 4.

The band structure and the FSs topology in models *1* and *2* show close similarity, except for splitting of some near-Fermi bands for model *2*. As a result, the Fermi surface for KFe$_2$AsSe (*2*) has five electron-like sheets in the corners of the Brillouin zone, four hole-like cylinders (along $\Gamma$-Z) and one closed electron-like pocket inside cylindrical-like sheets, Fig. 4. Anyhow, the aforementioned results indicate that KFe$_2$AsSe can exhibit superconductivity due to disconnected nested hole and electron Fermi surfaces [4-7].

Finally, the preliminary BCS-like estimations of $T_C$ for the proposed KFe$_2$AsSe can be performed, using the simplified model.[31] Here, the expression for $T_C = 1.14\omega_D e^{-2/gN(E_F)}$ was employed, and, as the authors [31] say, these estimations do not necessarily imply electron-phonon pairing, as $\omega_D$ may just denote the average frequency of any other possible Boson responsible for pairing interaction (e.g. spin fluctuations). Further, utilizing the average values of the Debye frequency $\omega_D$ = 350K, and the coupling constant $g$ = 0.21 eV, [31] and our calculated values of the total DOS at the Fermi level N(E$_F$) for KFe$_2$AsSe, rough values 22.6K for KFe$_2$AsSe (*1*) and 19.8K for KFe$_2$AsSe (*2*) were obtained, which appear comparable with $T_C$'s for known 122-like SCs. [1-7]

In summary, the ThCr$_2$Si$_2$-type arsenide-selenide phase KFe$_2$AsSe is proposed as a parent system for new "mixed" iron-pnictide-chalcogenide systems bridging the known families of Fe-pnictides (such as BaFe$_2$As$_2$) and Fe-chalcogenides (such as K$_x$Fe$_{2-y}$Se$_2$) superconductors. The characterization of the proposed charge-balanced phase by means of FLAPW-GGA approach covers the crystal structure, As/Se atomic ordering, stability, electronic bands, Fermi surface, and density of electronic states.

We show that the influence of As/Se distributions on the stability, structural and electronic properties of the KFe$_2$AsSe phase is very insignificant. Our calculations indicate that the electronic band structure and Fermi surface topology of the arsenide-selenide phase KFe$_2$AsSe is similar on the whole to those for other Fe-based SCs, but represents an interesting "intermediate" type between $A$Fe$_2$Pn$_2$ and $A'$Fe$_2$Ch$_2$ phases. Namely, for KFe$_2$AsSe two 2$D$-like bands above E$_F$ (along $\Gamma$-Z) form two cylindrical hole pockets (as in $A$Fe$_2$Pn$_2$ phases), two other bands below E$_F$ are responsible for the formation of the doubly degenerated electron pocket in the corners of the Brillouin zone (as in $A$Fe$_2$Pn$_2$ and $A'$Fe$_2$Ch$_2$ phases), and the fifth band intersects the Fermi level along $\Gamma$-Z forming the closed disconnected electron-like pocket (as in $A'Fe_2Ch_2$ phases), which is inserted into the hole cylindrical-like sheets. These results suggest that this system should be explored for possible superconductivity or magnetism associated with the Fermi surface nesting. In addition, since the $KFe_2AsSe$ phase exhibits structural indicators which are close to the "optimal" values, we assume that this system may be considered as a potentially favorable parent system for the search of new groups of "mixed" iron-pnictide-chalcogenide SCs. These data can serve as useful reference points to accelerate further experimental as well as theoretical efforts.

Another aspect is that the examined $KFe_2AsSe$ phase without cation deficiency, which strictly obeys the electroneutrality rule, calls for further studies. We speculate that a set of related phases $K_xFe_{2-y}As_{1-z}Se_z$ can be formed, where **cation deficiency** can be **controlled** by varying the As/Se ratio. This can provide an interesting platform for further theoretical and experimental investigations of the interplay of cation vacancies and magnetic and electronic properties of new Fe-based superconducting materials, which were launched recently using 122 Fe-$Ch$ systems as an example.[9-11,14-17,33-35] We speculate that the main interest in the magnetic phenomena for the proposed $KFe_2(As-Se)$ systems may be for the compositions ranging between examined charge-balanced quaternary phase $KFe_2AsSe$ and the ternary As-free phase $KFe_{2-x}Se_2$, *i.e.* for $KFe_{2-x}As_ySe_{2-y}$ where $1 < y < 0$. Indeed, for stoichiometric $KFe_2AsSe$ with completely filled Fe sheets, the magnetic ordering should be similar to that in other Fe-based phases, i.e. it should be antiferromagnetic (AFM) stripe-like collinear commensurate in-plane magnetic ordering.[1-7] On the other hand, for the ternary phase $KFe_{2-x}Se_2$ with long-range order of Fe-vacancies, an unique vacancy-induced block-antiferromagnetic state was found.[36] We believe that for the proposed compositions $KFe_{2-x}As_ySe_{2-y}$ ($1 < y < 0$) with variable amount of Fe-vacancies a ***rich variety of new magnetic ordering types*** can arise - as "indermediate" between the aforementioned simple stripe-like and vacancy-induced block-like AFM states ***controlled by cation deficiency***.

## ACKNOWLEDGMENTS

This work was supported by the Russian Foundation for Basic Research, Grants No. RFBR- 09-03-00946 and No. RFBR- 10-03-96008.

___________________________

**TABLE I.** Optimized lattice parameters (*a, b,* and *c*, in Å), so-called anion height ($\Delta z$, in Å), and As(Se)-Fe-As(Se) bond angles ($\Theta$, in degrees) for $KFe_2AsSe$ (*1 – 4*).

| Model* | Space group | Structural parameters | | | | |
| --- | --- | --- | --- | --- | --- | --- |
| | | *a* | *b* | *c* | $\Delta z$ | $\Theta$ ** |
| *1* | I4/*mmm* | 3.8252 | - | 13.7275 | 1.33 | 108.3-111.9 |
| *2* | P4/*mmm* | 3.8320 | - | 13.6374 | 1.33 | 108.4-112.1 |
| *3* | C/*mmm* | 5.4563 | 13.7950 | 5.4189 | 1.34 | 108.8-110.9 |
| *4* | C/*mmm* | 5.4345 | 13.7548 | 5.4204 | 1.34 | 108.9-110.8 |

* see Fig. 1.
** maximal and minimal angles are given

**TABLE II.** Total and partial densities of states at the Fermi level (in states/eV·f.u.) for $KFe_2AsSe$.

| Model* | *total* | Fe 3*d* | As 4*d* | Se 4*d* |
| --- | --- | --- | --- | --- |
| *1* | 3.32 | 2.72 | 0.04 | 0.03 |
| *2* | 3.16 | 2.59 | 0.04 | 0.03 |

* see Fig. 1.

**FIGURES**

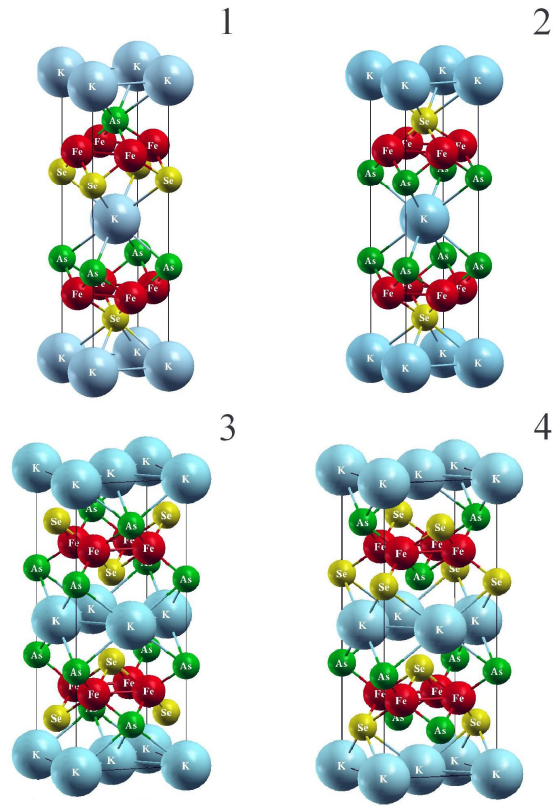

**FIG. 1.** (*Color online*) Structural models for the proposed KFe$_2$AsSe system with possible types of As/Se arrangement (*1-4*, see text).

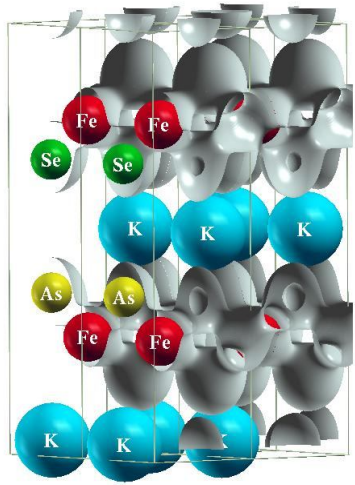

**FIG. 2.** (*Color online*) Charge density iso-surface of valence states for KFe$_2$AsSe system (model *1*, see Fig. 1).

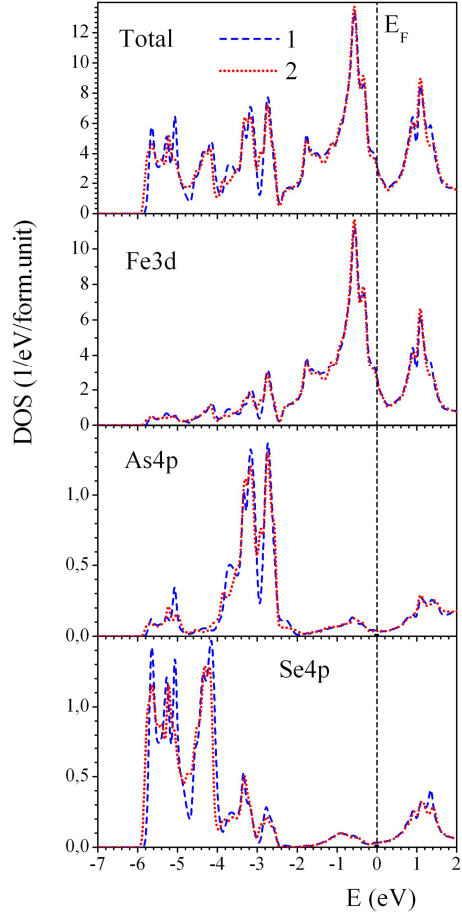

**FIG. 3.** (*Color online*) Total and partial densities of states of the KFe$_2$AsSe system for structural models *1* and *2*, see Fig. 1.

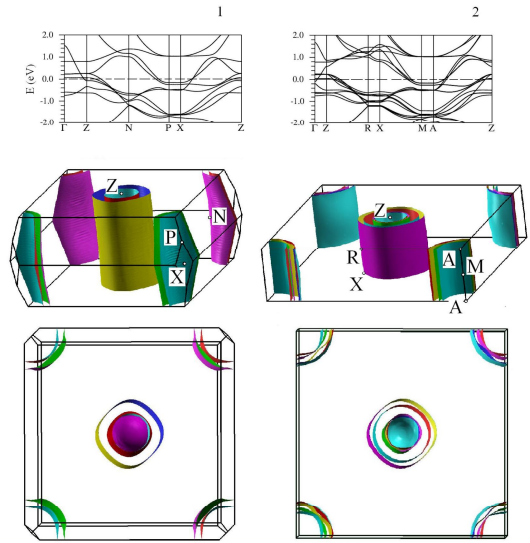

**FIG. 4.** (*Color online*) Electronic bands (top panels) and Fermi surfaces of the KFe$_2$AsSe system for structural models *1* and *2*, see Fig. 1. At the bottom panels, the closed disconnected electron-like pockets inserted into hole-like cylindrical sheets are shown.